\documentclass[%
 reprint,
 amsmath,amssymb,
 aps,
 prb,
floatfix,
]{revtex4-2}

\usepackage{xcolor}
\usepackage{graphicx}
\usepackage{dcolumn}
\usepackage{bm}
\usepackage{hyperref}

\begin{document}

\title{Ultrafast Photoexcitation of Semiconducting Photocathode Materials}

\author{Hilde Bellersen}
\email{hilde.bellersen@uni-oldenburg.de}
\author{Michele Guerrini}%
\email{michele.guerrini@uni-oldenburg.de}
\author{Caterina Cocchi}%
\email{caterina.cocchi@uni-oldenburg.de}
\affiliation{%
 Carl von Ossietzky Universit\"at Oldenburg, Institute of Physics, 26129 Oldenburg, Germany
}%

\date{\today}

\begin{abstract}
Cs-based semiconductors like $\mathrm{Cs_3Sb}$ and $\mathrm{Cs_2Te}$ are currently used as photocathodes in particle accelerators. Their performance as electron sources critically depends on their interaction with intense laser sources. In this work, we investigate from first principles the time-dependent response of $\mathrm{Cs_3Sb}$ and $\mathrm{Cs_2Te}$ to ultrafast pulses of varying intensities, ranging from $1~\mathrm{GW/cm^2}$ to $1~\mathrm{PW/cm^2}$. Nonlinear effects, including high harmonic generation, emerge starting from $100~\mathrm{GW/cm^2}$ in $\mathrm{Cs_3Sb}$ and $200~\mathrm{GW/cm^2}$ in $\mathrm{Cs_2Te}$. Above these intensities, the numbers of absorbed photons and excited electrons saturate due to the depletion of one-photon absorption channels, with renewed increases beyond $1~\mathrm{TW/cm^2}$ for $\mathrm{Cs_3Sb}$ and $5~\mathrm{TW/cm^2}$ for $\mathrm{Cs_2Te}$ where multi-photon absorption appears. Finally, the analysis of the occupation density in $\mathrm{Cs_3Sb}$ reveals the onset of tunnel ionization at intensities above $10~\mathrm{TW/cm^2}$. Our findings provide new insights into the nonlinear optical properties of $\mathrm{Cs_3Sb}$ and $\mathrm{Cs_2Te}$, contributing to the optimization of these materials for the development of next-generation photoinjectors.
\end{abstract}

\maketitle

\section{\label{sec:intro}Introduction}

Ultrabright electron beams generated by radiofrequency (RF) photoinjectors are an important tool in modern material science, offering a powerful way to explore in great detail the fundamental properties of different classes of compounds~\cite{tschentscher2017,sciaini2019}.
They are an essential part of advanced synchrotron facilities and free electron lasers around the world \cite{Emma2010,Dowell2010,Zhang2024}, where most of such experiments are performed. 
In RF photoinjectors, electrons are released from the photocathode through the photoelectric effect. The emitted electrons are then accelerated by strong electric fields, giving rise to ultrabright electron beams.
High-quality beams are achieved with high peak current for reduced gain length, low emittance for a small beam cross-section, and a narrow energy spread~\cite{Dowell2010, Ayvazyan2002, Alley1999}. 
Since the beam quality is directly linked to the properties of the photocathode, it is essential to select and optimize its composition and characteristics appropriately.
This need has driven extensive research in the last decade, focused on developing materials with high quantum efficiency, long durability, low thermal emittance, and fast response times~\cite{Dowell2010, Barday2013, Schmeier2018, Antoniuk2020, Loisch2022, Schaber2023}. 

While metals are the most established photocathode materials~\cite{Dowell2010,maldonado2008}, current trends in accelerator physics increasingly favor alkali-based semiconductors, thanks to their high quantum efficiency and ability to produce low-emittance beams at high repetition rates~\cite{Dowell2010,musumeci2018,Loisch2022}.
$\mathrm{Cs_2Te}$ and $\mathrm{Cs_3Sb}$ are among the most studied compounds, as they offer high quantum efficiency, good thermal stability, durability, and reliable performance~\cite{Filippetto2015, Liu2022, Parzyck2022}.
The rise of these materials has received essential support from computational studies based on \textit{ab initio} methods such as density-functional theory (DFT) and many-body perturbation theory, which have contributed to shed light on their fundamental properties, including their electronic structure,  vibrational properties, and response to radiation~\cite{Kalarasse2010, guo14mre, cocchi2019_2, Cocchi2019, Cocchi2020, amador2021, Cocchi2021, Sassnick_2021, schier2022, wu-gano23jmca, schier2024, SantanaAndreo2024, wang2025, xu2025}.

For a complete understanding of light-matter interactions in $\mathrm{Cs_2Te}$ and $\mathrm{Cs_3Sb}$, it is important to analyze their (non)linear response to ultrafast coherent pulses of varying intensities approaching those impinging the photocathodes to produce the electron beams. The most suitable \textit{ab initio} scheme for this purpose is real-time time-dependent DFT (RT-TDDFT)~\cite{bertsch2000}, which provides a reliable description of the evolution of the electron density subject to an external time-dependent electric field of tunable frequency, polarization, shape, duration, and intensity. Being a non-perturbative approach, RT-TDDFT delivers the complete response of the material including nonlinear effects~\cite{mait16jcp}. While initially implemented for confined systems like molecules and clusters~\cite{yabana1999}, this formalism has been successfully applied to several classes of crystalline materials, ranging from bulk semiconductors~\cite{Otobe2008, zhang2020, Cabrera2024} and insulators~\cite{wang2012, kong2018} to monolayers~\cite{su2017, tancogne2018, Uemoto2021} and their heterostructures~\cite{zhang2017, iida2018, Jacobs2022}.

In this work, we study from first principles the excited-state dynamics of $\mathrm{Cs_2Te}$ and $\mathrm{Cs_3Sb}$ crystals impinged by femtosecond pulses in resonance with their dominant absorption peaks in the ultraviolet (UV) and visible region. By irradiating the systems with intensities ranging from 1~GW/cm$^2$ up to 1~PW/cm$^2$, we investigate optical nonlinearities such as high-harmonic generation and multi-photon absorption. By analyzing critical quantities such as the numbers of excited electrons and absorbed photons, we estimate the absorption efficiency as a function of the laser intensity, identifying thresholds above which the response of the system is dominated by nonlinear effects. This knowledge is essential to optimize the performance of semiconducting photocathodes for ultrabright electron beams.


\section{\label{sec:methodology}Computational methods}

\subsection{Theoretical Background}

The calculations presented in this study are based on DFT~\cite{dft,KS} and RT-TDDFT~\cite{yabana1996} as implemented in the \textsc{Octopus} code~\cite{TancogneDejean2020} adopting a real-space grid representation for the electron density. 
Linear absorption spectra are computed by perturbing the systems with an instantaneous broad-band electric field ($\delta$-kick) in the velocity gauge~\cite{bertsch2000}. The optical conductivity $\sigma(\omega)$ is obtained by dividing the current density by the incident electric field $\mathcal{E}(\omega)$ induced by the $\delta$-kick:
\begin{equation}
    \sigma(\omega) = \frac{J(\omega)}{\mathcal{E}(\omega)}.
    \label{eq:sigma}
\end{equation}
The imaginary part of the dielectric function $\varepsilon(\omega)$ is related to the real part of the optical conductivity by the conventional formula:
\begin{equation} \label{Re_Im_conv}
  \Im [\varepsilon (\omega )] = \frac{4 \pi \Omega}{\mathcal{E}_0 \omega} \Re [\sigma (\omega)],
\end{equation}
where $\omega$ is the frequency of the incoming radiation, $\Omega$ the unit-cell volume, and $\mathcal{E}_0$ the amplitude of $\mathcal{E}(\omega)$.

The time-dependent response of the system under an external ultrafast electric field is evaluated by applying a vector potential~\cite{yabana2006} connected to the electric field by the usual relation
\begin{equation}
    \mathbf{\mathcal{E}}(t) = - \frac{d\mathbf{A}(t)}{dt}.
\end{equation}
The microscopic current density 
\begin{equation}
    \mathbf{j}(\mathbf{r},t) = \frac{1}{2N_k} \sum_{n\mathbf{k}} \left[ \phi^*_{n\mathbf{k}}(\mathbf{r},t) \left(-i\nabla + \frac {\mathbf{A}(t)}{c}\right) \phi_{n\mathbf{k}}(\mathbf{r},t) + c.c.
    \right]
    \label{eq:j}
\end{equation}
is obtained from the solutions of the time-dependent Kohn-Sham (KS) equations, $\phi_{n\mathbf{k}}(\mathbf{r},t)$, where $N_k$ is the total number of k-points sampling the Brillouin zone.
Integrating Eq.~\eqref{eq:j} in the unit cell volume $\Omega$ leads to the time-dependent macroscopic current density
\begin{equation}
    \mathbf{J}(t) = \frac{1}{\Omega} \int_\Omega 
                    \mathbf{j}(\mathbf{r},t) d^3 r ,
\end{equation}
which enters the expression of the high harmonic generation (HHG) spectrum: 
\begin{equation}
\label{eq:J_omega}
    H(\omega) = \left| \int_0^\infty \frac{\mathbf{J}(t)}{dt}e^{-i\omega t} dt \right|^2.
\end{equation}

The population of the time-dependent KS states is obtained from their projection onto the KS states in the ground state $(t=0)$:
\begin{equation}
    \mathcal{P}_{n,\textbf{k}}(t)=\sum_{n'}^{occ} |\langle \phi_{n\textbf{k}}(t=0)|\phi_{n'\textbf{k}}(t)\rangle|^2.
\end{equation}
The occupation density is given by the ground-state density of states weighted by  $\mathcal{P}_{n,\textbf{k}}(t)$:
\begin{equation}
    \mathcal{N}(\epsilon,t) = \sum_{n,\textbf{k}}^{all} \delta (\epsilon - \epsilon_{n,\textbf{k}}^0) \mathcal{P}_{n,\textbf{k}}(t),
\end{equation}
while the number of excited electrons per unit cell is obtained by summing $\mathcal{P}_{n,\textbf{k}}(t)$ over all unoccupied states:
\begin{equation}
    N_{ex}(t) = \sum_{n,\mathbf{k}}^{unocc} \mathcal{P}_{n,\mathbf{k}}(t).
    \label{eq:nex}
\end{equation}
The time-dependent energy per unit volume absorbed by the system from the field  corresponds to the difference between the total energy  at a certain time $t$ after the perturbation and the ground state energy, $\Delta E(t) = E(t) - E_{GS}$. This quantity is related to the total current and the external field via
\begin{equation}
    \Delta E(t)= \int_0^t \textbf{J}(t') \cdot \mathcal{E}(t') dt'.
\end{equation}
The number of photons absorbed by the irradiated material is the ratio between $\Delta E$, evaluated at the end of the pulse propagation ($t=t_{fin}$), and the energy associated with the carrier frequency of the applied field ($\hbar\omega_0$): 
\begin{equation}
\label{eq:N_ph}
    N_{ph}=\dfrac{\Delta E(t_{fin})}{\hbar\omega_0}.
\end{equation}
The number of incoming photons is estimated by calculating the total energy of the pulse $E_{pulse}$, obtained by integrating the intensity over time and multiplying it by the effective unit cell area $A$, divided by the photon energy $E_{photon}=\hbar\omega_0$,
\begin{equation}
    N_{ph}^{in}= \frac{E_{pulse}}{E_{photon}} = \frac{I \!\cdot\!\sqrt{2\pi}\!\cdot\!\tau_0 \cdot A}{\hbar \omega_0},
\end{equation}
where $I$ is the peak intensity of the pulse and $\tau_0$ the standard deviation of the Gaussian envelope. 
The photon absorption efficiency is finally defined as the ratio between the numbers of absorbed and incident photons:
\begin{equation}
\label{eq:photon_eff}
    \mathcal{A}_{ph} = \dfrac{N_{ph}}{N_{ph}^{in}}.
\end{equation}

\subsection{Computational Details}

The ground-state calculations were carried out on a real-space grid with 0.2~\AA{} spacing and k-meshes with 5$\times$5$\times$5 points for $\mathrm{Cs_3Sb}$ and 4$\times$6$\times$4 points for $\mathrm{Cs_2Te}$, obtained from a modified version of the Monkhorst-Pack scheme~\cite{Monkhorst1976}. Norm-conserving ONCV pseudopotentials~\cite{Schlipf2015} are used to treat core electrons and the Perdew-Burke-Ernzerhof (PBE) approximation~\cite{pbe} is adopted for the exchange-correlation potential in DFT and for the exchange-correlation functional in RT-TDDFT. We did not include spin-orbit coupling based on previous findings~\cite{Sassnick_2021} indicating a negligible role of this effect on the band structures of both $\mathrm{Cs_3Sb}$ and $\mathrm{Cs_2Te}$. The lattice parameters of both materials are taken from Ref.~\cite{Sassnick_2021} without performing any further optimization.

In the RT-TDDFT runs, the systems are perturbed with a $x$-polarized pulse of Gaussian shape. For $\mathrm{Cs_3Sb}$,  the peak intensity was set at $t_0=$~18~fs with a standard deviation $\tau_0=$~5~fs, while for $\mathrm{Cs_2Te}$, $t_0=15$~fs and $\tau_0=4$~fs. The pulse carrier frequency was set to 2.4~eV/$\hbar$ for $\mathrm{Cs_3Sb}$ and 3.6~eV/$\hbar$ for $\mathrm{Cs_2Te}$, in resonance with the absorption maxima dominating the linear spectra of the two materials, to ensure that the incoming energy is efficiently uptaken. Pulses of increasing intensities are adopted, ranging from $1~\mathrm{GW/cm^2}$ to $1~\mathrm{PW/cm^2}$, with a sufficient number of intermediate steps to accurately probe the nonlinear response of $\mathrm{Cs_3Sb}$ and $\mathrm{Cs_2Te}$. The time propagations were run for a total duration of 70~fs for $\mathrm{Cs_3Sb}$ and 60~fs for $\mathrm{Cs_2Te}$, with time steps ranging from 0.7~as to 2.2~as depending on the laser intensity.
To minimize the numerical divergence at low-frequency of the linear absorption spectra obtained with the $\delta$-kick method, we increased the density of the k-grids to 10$\times$10$\times$10 points for $\mathrm{Cs_3Sb}$ and 5$\times$8$\times$4 points for $\mathrm{Cs_2Te}$. For these calculations, the propagation time was set to 10~fs with time steps of 1~as for $\mathrm{Cs_3Sb}$ and 0.5~as for $\mathrm{Cs_2Te}$.

\section{\label{sec:results}Results and Discussion}

\subsection{Structural Properties}

The crystal structure of $\mathrm{Cs_3Sb}$ consists of a face-centered cubic (fcc) lattice with space group $Fm\overline{3}m$, already adopted in previous theoretical studies \cite{Sassnick_2021, Cocchi2019, Wei1987, Kalarasse2010, Cocchi2020} as a simplification of the cubic lattice originally proposed by~\citet{Jack1957}.
The primitive unit cell, shown in the inset of Fig.~\ref{fig:absorption}a, includes three Cs atoms at fractional positions ($\frac{1}{4}$, $\frac{1}{4}$, $\frac{1}{4}$), ($\frac{1}{2}$, $\frac{1}{2}$, $\frac{1}{2}$), and ($\frac{3}{4}$, $\frac{3}{4}$, $\frac{3}{4}$) with the first and last being chemically equivalent~\cite{Cocchi2020}, and one Sb atom located at the origin.
The adopted lattice parameter of 9.298~\AA{} is taken from the PBE results of Ref.~\cite{Sassnick_2021}. This value overestimates the experimental reference by $1.18\%$, a known shortcoming of the PBE functional \cite{Zhang2018}.

$\mathrm{Cs_2Te}$ belongs to the $Pnma$ space group and is characterized by an orthorhombic unit cell with lattice vectors $a=$~9.542~\AA{}, $b=$~5.845~\AA{}, $c =$~11.591~\AA{} taken from the PBE results of Ref.~\cite{Sassnick_2021}. The unit cell, depicted in the inset of Fig.~\ref{fig:absorption}b, includes eight Cs atoms and four Te atoms, and corresponds to the experimentally established polymorph of this material~\cite{schewe1991}. Other phases of $\mathrm{Cs_2Te}$, which have been computationally predicted to be stable~\cite{sassnick2022}, are not considered in this work. The band structures computed for both materials are reported in the Supplemental Material, Fig.~S1.


\subsection{Linear Absorption Spectra}

\begin{figure}
    \centering
    \includegraphics[width=1\linewidth]{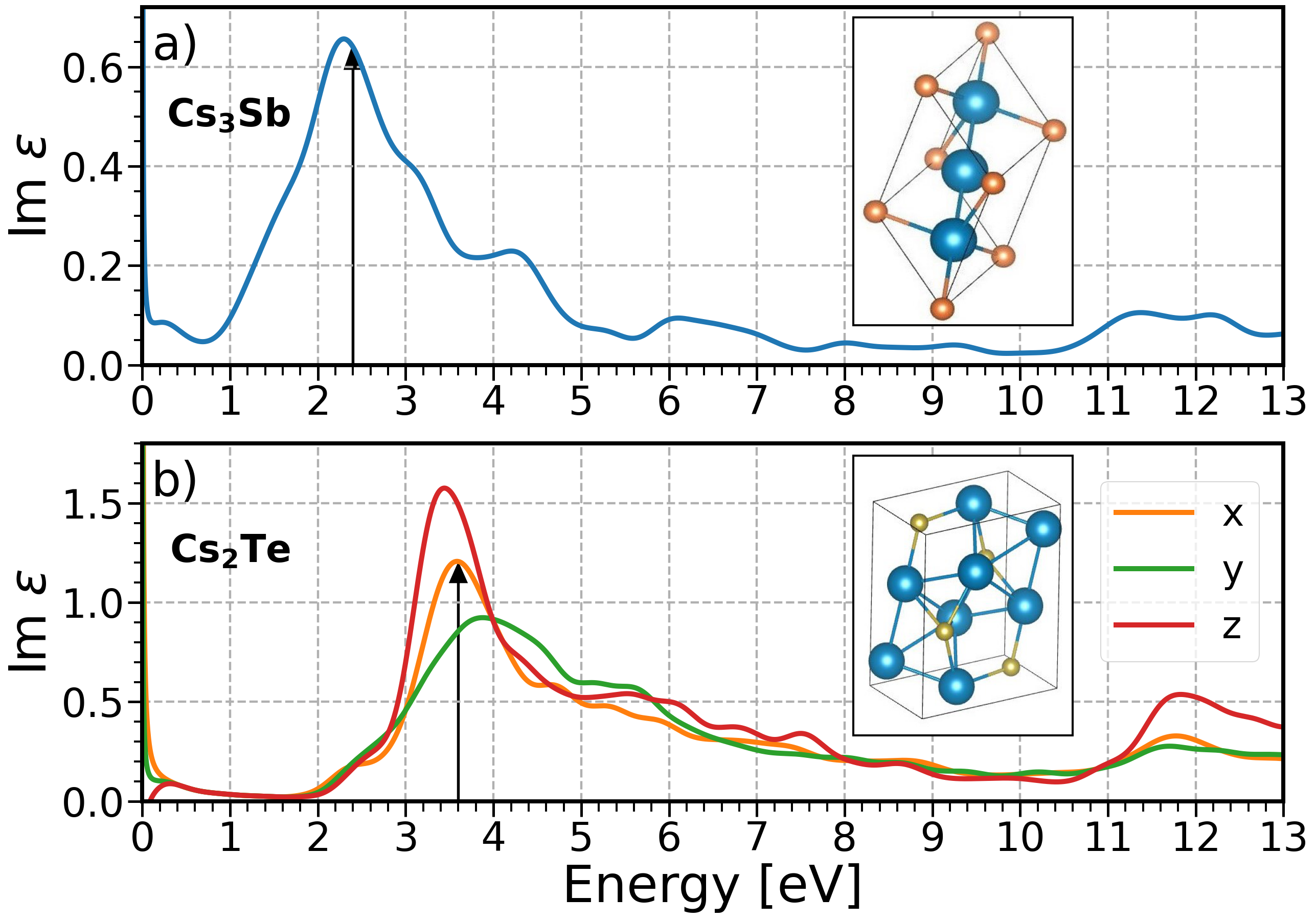}
    \caption{Linear absorption spectra of a) $\mathrm{Cs_3Sb}$ and b) $\mathrm{Cs_2Te}$ calculated as the imaginary part of their dielectric function. In contrast to cubic $\mathrm{Cs_3Sb}$, characterized by a single inequivalent component of $\varepsilon$, the spectrum of orthorhombic $\mathrm{Cs_2Te}$ exhibits distinct contributions along each Cartesian polarization direction. 
Insets: Ball-and-stick representations, generated with the visualization software VESTA~\cite{Momma2011}, of the unit cells of a) $\mathrm{Cs_3Sb}$ and b) $\mathrm{Cs_2Te}$, with Cs atoms depicted in blue, Sb atoms in orange, and Te atoms in gold.}
    \label{fig:absorption}
\end{figure}

The linear absorption spectrum of $\mathrm{Cs_3Sb}$ is dominated by a strong peak at 2.4~eV preceded by a shoulder around 1.5~eV (Fig.~\ref{fig:absorption}a). Weaker maxima are found around 3~eV and 4.5~eV while deeper in the UV region, the material is almost transparent to radiation. The main absorption features displayed in Fig.~\ref{fig:absorption}a are consistent with those obtained from the $GW$ approximation and the solution of the Bethe-Salpeter equation (BSE)~\cite{Cocchi2021}, likely due to a mutual compensation of quasi-particle corrections and exciton binding energies.

The absorption spectrum of $\mathrm{Cs_2Te}$ (Fig.~\ref{fig:absorption}b) starts at approximately 2~eV and exhibits sharp maxima at 3.6~eV in the $x$-polarization and at 3.45~eV along $z$. In the $y$-direction, a broader peak appears at 3.9~eV. Similar to $\mathrm{Cs_3Sb}$, also $\mathrm{Cs_2Te}$ is almost transparent to near-UV light except around 12~eV, where a weak maximum appears especially in the $z$-component. Due to larger and not matching values of quasi-particle corrections and exciton binding energies in this material, the agreement between Fig.~\ref{fig:absorption}b and the $GW$/BSE results reported in Ref.~\cite{Cocchi2021} is worse than for $\mathrm{Cs_3Sb}$. Nonetheless, the qualitative characteristics of the spectrum, with weak absorption in the visible and intense peaks at the boundary with the UV region, are robust with respect to the adopted formalism.


\subsection{High Harmonic Generation}

\begin{figure*}
    \centering
    \includegraphics[width=0.9\linewidth]{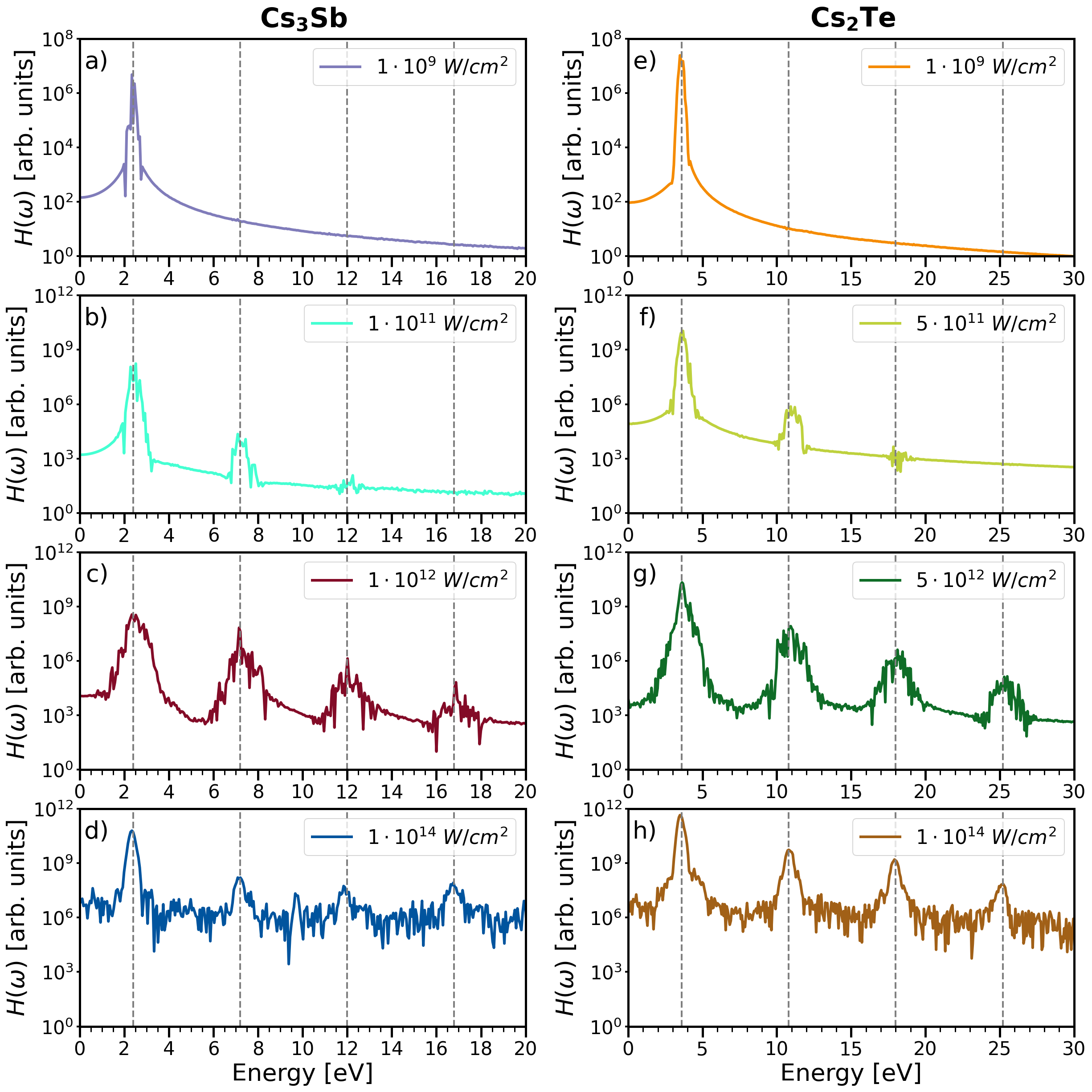}
    \caption{HHG spectra of $\mathrm{Cs_3Sb}$ (left) and $\mathrm{Cs_2Te}$ (right) impinged by pulses with intensity 
    a) $I=1~\mathrm{GW/cm^2}$,  
    b) $I=100~\mathrm{GW/cm^2}$, 
    c) $I=1~\mathrm{TW/cm^2}$,
    d) $I=100~\mathrm{TW/cm^2}$,
    e) $I=1~\mathrm{GW/cm^2}$,  
    f) $I=500~\mathrm{GW/cm^2}$, 
    g) $I=5~\mathrm{TW/cm^2}$,
    h) $I=100~\mathrm{TW/cm^2}$.
    The vertical dashed bars mark the energies of the fundamental harmonics and their odd multiples.
    }
    \label{fig:HHG}
\end{figure*}

With the knowledge of the linear optical properties of $\mathrm{Cs_3Sb}$ and $\mathrm{Cs_2Te}$, we proceed with the analysis of their time-dependent response to an ultrafast electric field set in resonance with their most intense excitation in the UV-visible region (black arrows in Fig.~\ref{fig:absorption}).
In the case of $\mathrm{Cs_2Te}$, we excite the system along the $x$-direction, thus targeting the maximum at 3.6~eV.
We first examine HHG spectra (Eq.~\ref{eq:J_omega}), comparing for both materials the results obtained with four increasing laser intensities ranging from $1~\mathrm{GW/cm^2}$ to $100~\mathrm{TW/cm^2}$, see Fig.~\ref{fig:HHG}. 

With the weakest pulse ($I=1~\mathrm{GW/cm^2}$), the HHG spectrum of $\mathrm{Cs_3Sb}$ contains only the fundamental harmonic at $\hbar\omega_0=2.4~\mathrm{eV}$, corresponding to the carrier frequency of the pulse (Fig.~\ref{fig:HHG}a). In this linear regime, the response of the system can be described by the microscopic form of Ohm's law, where the current density is proportional to the applied electric field
\begin{equation}
\label{eq: Ohms_law}
    J(\omega) = \sigma(\omega) \mathcal{E}(\omega),
\end{equation}
which is consistent with Eq.~\eqref{eq:sigma}.
As shown in Fig.~\ref{fig:HHG}b, by increasing the laser intensity to $100~\mathrm{GW/cm^2}$, higher harmonics are excited. Due to the inversion symmetry of the $\mathrm{Cs_3Sb}$ crystal, the second harmonic is silent while the third one is clearly visible at $3\hbar\omega_0=7.2~\mathrm{eV}$. Further increase of the field intensity to 100~$\mathrm{GW/cm^2}$ excites also the fifth harmonic, generating a weak but distinct signal at $5\hbar\omega_0=12~\mathrm{eV}$. These findings indicate a nonlinear response of the system, where the current density is no longer proportional to the applied electric field as in Eq.~\eqref{eq: Ohms_law}. Instead, the generation of higher harmonics indicates the appearance of nonlinear optical processes, involving both interband and intraband transitions~\cite{Golde2008, Vampa2014}.

By further increasing the laser intensity to $1~\mathrm{TW/cm^2}$, the magnitudes of the third and fifth harmonic increase, and  higher harmonics up to the eleventh order are generated (see Supplemental Material, Fig.~S2). In the energy range visualized in Fig.~\ref{fig:HHG}c, the seventh harmonic is visible at $7\hbar\omega_0=16.8~\mathrm{eV}$ in addition to the third and fifth ones.
When the external field reaches the intensity of $100~\mathrm{TW/cm^2}$, the baseline of the HHG spectrum is significantly elevated across the entire frequency range shown in Fig.~\ref{fig:HHG}d. In this case, the harmonic peaks become much less distinct due to strong-field effects, including enhanced ionization and tunneling \cite{Vampa2014}. The non-vanishing spectral components at frequencies corresponding to even high harmonics hint at a (dynamical) symmetry breaking in the density distribution of $\mathrm{Cs_3Sb}$ induced by the strong electric field~\cite{Driscoll1983}, along with other complex nonlinear effects that cannot be straightforwardly singled out in the adopted non-perturbative approach.

The results obtained for $\mathrm{Cs_2Te}$, displayed on the right side of Fig.~\ref{fig:HHG}, are qualitatively similar to those of $\mathrm{Cs_3Sb}$ discussed above.
The weakest pulse intensity of $1~\mathrm{GW/cm^2}$ induces only the fundamental harmonic at $\hbar\omega_0=$~3.6~eV (Fig.~\ref{fig:HHG}e).
The third-harmonic signal (we recall that also $\mathrm{Cs_2Te}$ is centrosymmetric and thus even harmonics are forbidden) at $3\hbar\omega_0=$~10.8~eV is triggered by a laser intensity of $500~\mathrm{GW/cm^2}$ (Fig.~\ref{fig:HHG}f), which is a factor 5 larger compared to the one needed to excite the same component in $\mathrm{Cs_3Sb}$ (Fig.~\ref{fig:HHG}b). An intensity of 5~$\mathrm{TW/cm^2}$ activates higher harmonics up to $7\hbar\omega_0=25.2~\mathrm{eV}$, see Fig.~\ref{fig:HHG}g. The HHG signal generated by $I=100~\mathrm{TW/cm^2}$ shows again an elevated baseline as in $\mathrm{Cs_3Sb}$ (compare Fig.~\ref{fig:HHG}h and Fig.~\ref{fig:HHG}d). However, in contrast to $\mathrm{Cs_3Sb}$, in the HHG spectrum of $\mathrm{Cs_2Te}$, the peaks associated with the odd harmonics remain distinct from the background noise and minima appear at the energies of the even harmonics, suggesting a higher threshold for tunneling and ionization for this material.

Overall, the results presented in Fig.~\ref{fig:HHG} indicate a higher HHG yield in $\mathrm{Cs_2Te}$ than in $\mathrm{Cs_3Sb}$. This enhancement can be attributed to two main factors. Firstly, the slightly higher atomic number of Te (Z=52) compared to Sb (Z=51) leads to a stronger electron-nuclear potential due to  increased Coulomb interaction~\cite{TancogneDejean2017}. Although subtle, this effect cannot be ignored. Secondly, the lower proportion of Cs atoms with respect to the non-metallic species in the unit cell of $\mathrm{Cs_2Te}$ compared to  $\mathrm{Cs_3Sb}$ reduces the influence of delocalized Cs electrons, resulting in a greater contribution from the heavier Te atoms to the electron-nuclear potential, leading to a more pronounced spatial variation of the electron density.


\subsection{Number of Absorbed Photons, Number of Excited Electrons, and Photon Absorption Efficiency}

\begin{figure}
    \centering
    \includegraphics[width=\linewidth]{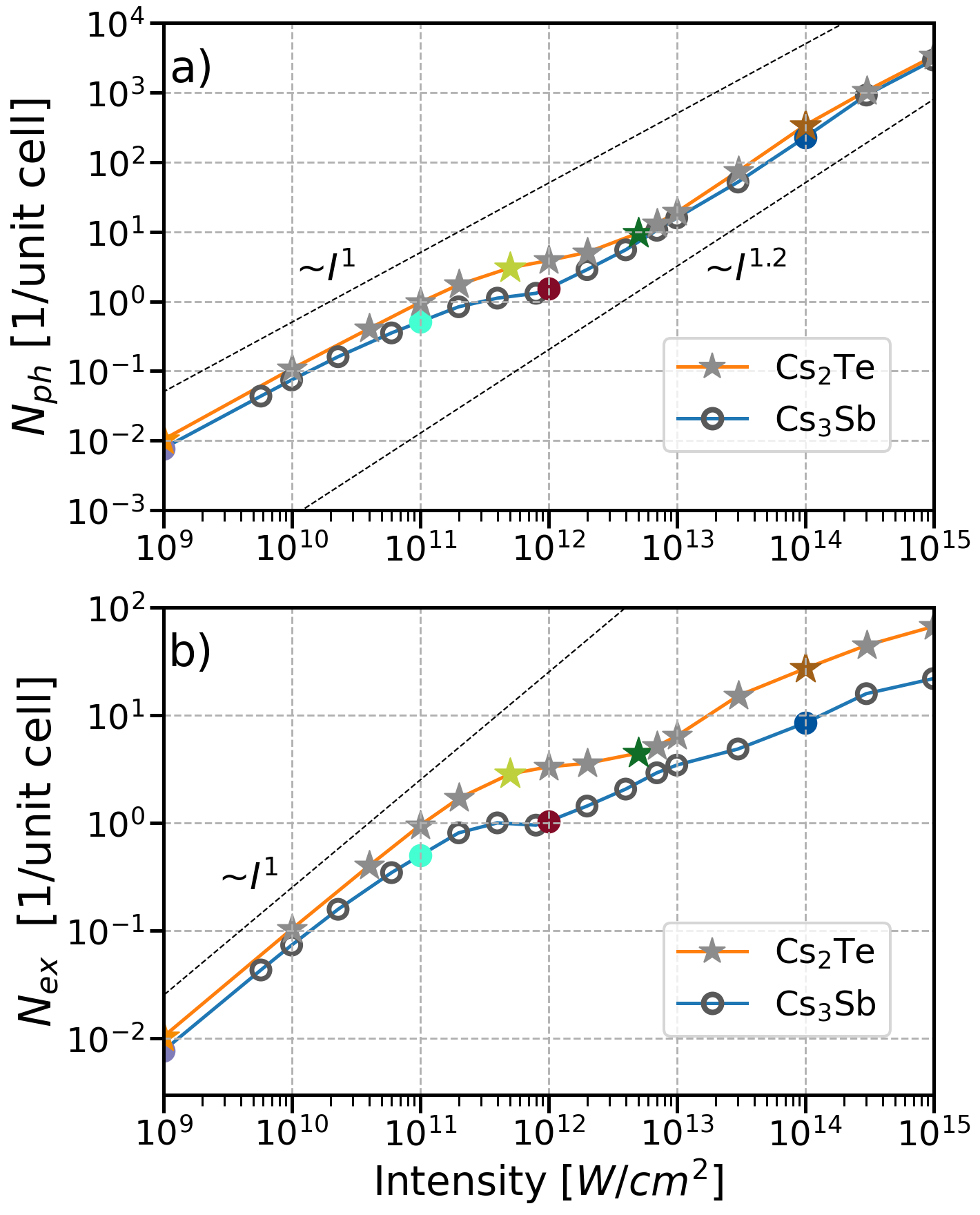}
\caption{a) Number of absorbed photons $N_{ph}$ and b) number of excited electrons $N_{ex}$ as a function of the laser intensity. Both quantities are evaluated at $t=36$~fs when the pulse is switched off. Color-filled symbols are used for the intensities adopted to plot the HHG spectra (same color code used) while gray stars and hollow circles correspond to additional calculations performed to evaluate the overall trend. The solid colored lines are guides for the eyes, while the dashed lines indicate linear trends proportional to the indicated power of the laser intensity $I$.}
    \label{fig:E_N_both}
\end{figure}

To further assess the nonlinear response of $\mathrm{Cs_3Sb}$ and $\mathrm{Cs_2Te}$ to resonant pulses of increasing intensity, we calculate macroscopic quantities such as the number of absorbed photons, the number of excited electrons, and the photon absorption efficiency per unit volume of each material, all evaluated after the pulse is turned off. The temporal evolution of $N_{ph}(t)$ and $N_{ex}(t)$ is shown in the Supplemental Material, Fig.~S3 and Fig.~S4.

By inspecting Fig.~\ref{fig:E_N_both}a, where the number of absorbed photons are displayed as a function of the laser intensity, we notice that in both $\mathrm{Cs_3Sb}$ and $\mathrm{Cs_2Te}$, $N_{ph}$ increases linearly up to $I=100~\mathrm{GW/cm^2}$ in $\mathrm{Cs_3Sb}$ and to $I=200~\mathrm{GW/cm^2}$ in $\mathrm{Cs_2Te}$. Beyond these values, $N_{ph}$ reaches a plateau, suggesting the saturation onset for one-photon absorption, while multi-photon absorption remains negligible. At $I>1~\mathrm{TW/cm^2}$ for $\mathrm{Cs_3Sb}$ and $I> 5~\mathrm{TW/cm^2}$ for $\mathrm{Cs_2Te}$, weak multi-photon absorption starts to appear, as confirmed by the superlinear increase of the number of absorbed photons, $N_{ph}(I) \propto I^{1.2}$. The trends obtained for the two materials are quite similar although in $\mathrm{Cs_2Te}$, $N_{ph}$ remains systematically higher than in $\mathrm{Cs_3Sb}$.

The number of excited electrons displayed in Fig.~\ref{fig:E_N_both}b exhibits a more complex behavior. A linear increase proportional to the pulse intensity appears up to $100~\mathrm{GW/cm^2}$ for $\mathrm{Cs_3Sb}$ and $200~\mathrm{GW/cm^2}$ $\mathrm{Cs_2Te}$, where $N_{ph}(I) \propto I$. Above these thresholds, $N_{ex}$ forms a plateau similarly to $N_{ph}$, suggesting a saturation of the lowest unoccupied states due to Pauli blocking~\cite{zhang2020,Jacobs2022}. For $I>1~\mathrm{TW/cm^2}$ and $I > 5~\mathrm{TW/cm^2}$, when multi-photon absorption comes into play, the number of excited electrons starts to increase linearly again due to excited state absorption. Given the limited number of available electrons to be excited in this regime, $N_{ex}$ grows less steeply than at weaker intensities, where absorption from the ground state dominates. Eventually, $N_{ex}$ begins to saturate again at higher intensities, due to the finite number of electrons available in the unit cell.
In Fig.~\ref{fig:E_N_both}b, the number of excited electrons remains systematically larger for $\mathrm{Cs_2Te}$ than for $\mathrm{Cs_3Sb}$, although the discrepancy between the two materials increases at higher intensities. This behavior is consistent with the fact that $N_{ex}$ is directly related to the occupation density, see Eq.~\eqref{eq:nex}.

\begin{figure}
    \centering
    \includegraphics[width=\linewidth]{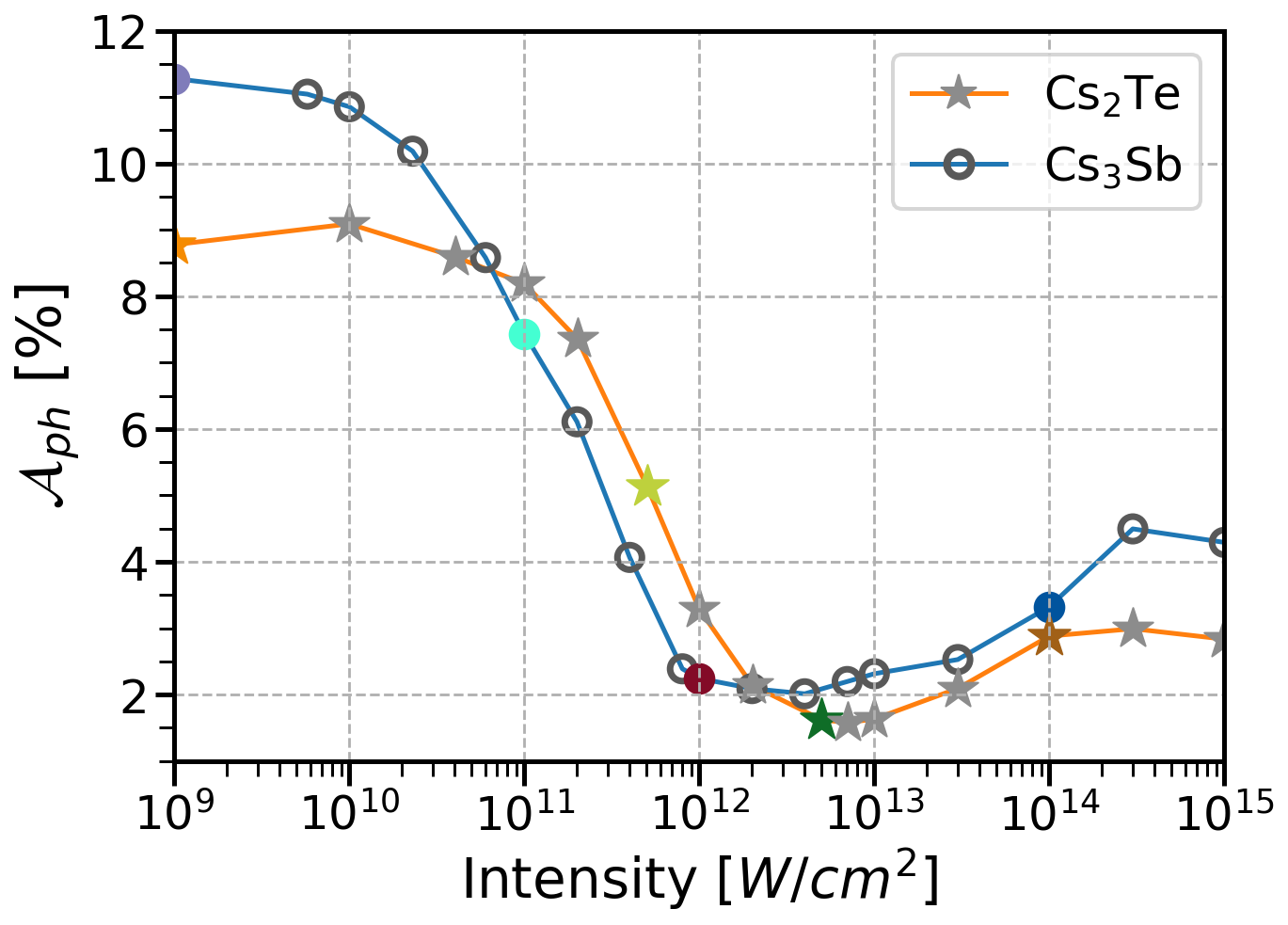}
    \caption{Photon absorption efficiency $\mathcal{A}_{ph}$ as a function of the laser intensity in $\mathrm{Cs_2Te}$ (orange) and $\mathrm{Cs_3Sb}$ (blue). Color-filled symbols are used for the intensities adopted to plot the HHG spectra (same color code used) while gray stars and hollow circles correspond to additional calculations performed to evaluate the overall trend.}
    \label{fig:abs_eff}
\end{figure}

We finally inspect the intensity-dependent photon absorption efficiency given by the ratio between absorbed and incoming photons (Eq.~\ref{eq:photon_eff}).
To perform this estimate, we take the effective area $A$, entering Eq.~\eqref{eq:photon_eff}, as the cross-sectional area of the unit cell of each crystal normal to the $x$-polarized incoming laser. The resulting values are 22~\AA{}$^2$ for $\mathrm{Cs_3Sb}$ and 67.7~\AA{}$^2$ for $\mathrm{Cs_2Te}$. 
As shown in Fig.~\ref{fig:abs_eff}, $\mathcal{A}_{ph}$ exhibits a similar behavior for both materials, starting with approximately constant (or, for $\mathrm{Cs_2Te}$, even slightly increasing) values under weak intensities ($I \leq 10~\mathrm{GW/cm^2}$), followed by a monotonic but nonlinear decrease up to $I = 1~\mathrm{TW/cm^2}$ associated to the region of saturated absorption. The absorption efficiency of $\mathrm{Cs_3Sb}$ is initially higher than that of $\mathrm{Cs_2Te}$ for weak lasers, but starts to decrease earlier, leading to a crossover at $50~\mathrm{GW/cm^2}$. At $I = 2~\mathrm{TW/cm^2}$, the absorption efficiency of $\mathrm{Cs_3Sb}$ again surpasses the one of $\mathrm{Cs_2Te}$ and remains above it for all higher intensities. Local minima appear at approximately $4~\mathrm{TW/cm^2}$ in $\mathrm{Cs_3Sb}$ and $7~\mathrm{TW/cm^2}$ in $\mathrm{Cs_2Te}$. In $\mathrm{Cs_3Sb}$, the photon absorption efficiency is almost flat between $I = 1~\mathrm{TW/cm^2}$ and $I = 10~\mathrm{TW/cm^2}$. Above this value, it increases again in both materials, before forming another plateau between $I = 100~\mathrm{TW/cm^2}$ and $I = 1~\mathrm{PW/cm^2}$.

We can correlate this behavior with the trends collected for the HHG spectra, $N_{ex}$, and $N_{ph}$. In the weak intensity range, where the response of the materials is linear ($I \leq 10~\mathrm{GW/cm^2}$), the photon absorption efficiency is maximized. Its steepest decrease appears for intensities triggering nonlinearities ($10~\mathrm{GW/cm^2} < I \leq 1~\mathrm{TW/cm^2}$) but where one-photon absorption still dominates. At the threshold of the strong coupling regime, where multi-photon processes become relevant ($I \sim 10~\mathrm{TW/cm^2}$), $\mathcal{A}_{ph}$ is minimized, due to the reorganization of the electronic structure. Multi-photon processes enable a further increase of the photon absorption efficiency between $I = 10~\mathrm{TW/cm^2}$ and $I = 100~\mathrm{TW/cm^2}$, where, however, the electronic structure of the material is irreversibly perturbed by the incoming radiation, as seen in the HHG spectra (Fig.~\ref{fig:HHG}d and h). This effect is more pronunced in $\mathrm{Cs_3Sb}$ than in $\mathrm{Cs_2Te}$.


\subsection{Occupation Density}
\label{ssec:occ_dens}

\begin{figure*}
    \centering
    \includegraphics[width=0.9\linewidth]{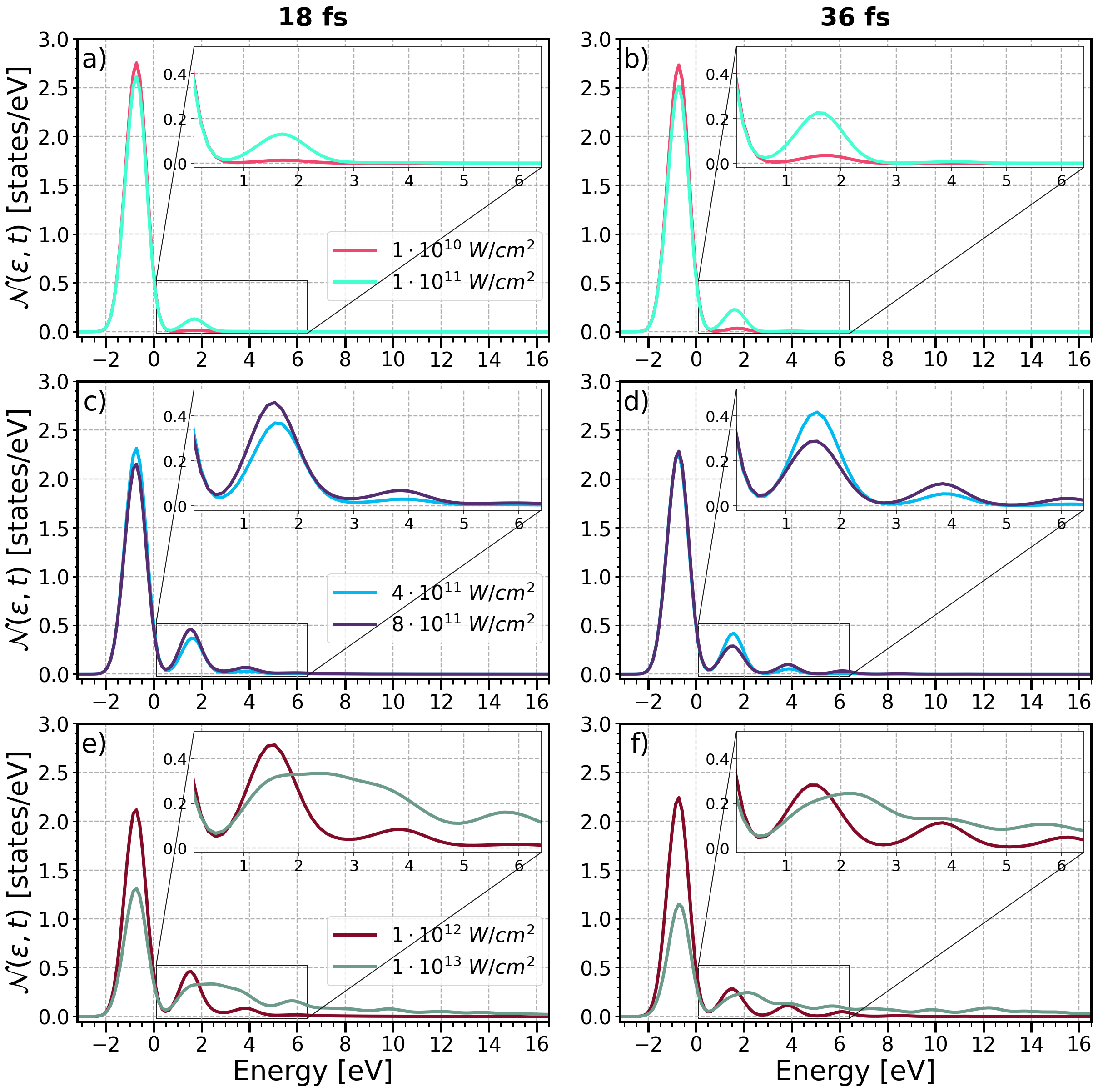}
    \caption{Occupation density $\mathcal{N}(\epsilon,t)$ of $\mathrm{Cs_3Sb}$ computed at the peak of the incident pulse ($t=18$~fs, left) and at the end of its propagation ($t=36$~fs, right) with intensities 
    a), b)~$I=10~\mathrm{GW/cm^2}$ and $I=100~\mathrm{GW/cm^2}$, 
    c), d)~$I=400~\mathrm{GW/cm^2}$ and $I=800~\mathrm{GW/cm^2}$, and 
    e), f)~$I=1~\mathrm{TW/cm^2}$ and $I=10~\mathrm{TW/cm^2}$.
    A Gaussian broadening of 38~meV is applied to all curves to enhance visualization.}
    \label{fig:occ_density}
\end{figure*}

We finally inspect the occupation density of $\mathrm{Cs_3Sb}$ evaluated at two relevant times, namely when the pulse is at its peak value (18~fs) and when it is turned off (36~fs). Given the similarities between the two materials, we focus only on $\mathrm{Cs_3Sb}$. The goal of this analysis is to gain insight into how the laser-excited electronic states redistribute over the unoccupied bands at increasing intensity of the incoming radiation, representing the regimes of weak, intermediate, and strong light-matter couplings. 

The interaction with a pulse stronger than $10~\mathrm{GW/cm^2}$ leads to a very small increase ($\sim$0.01 with $I=10~\mathrm{GW/cm^2}$ and $\sim$0.1 with $I=100~\mathrm{GW/cm^2}$) in the population of unoccupied states at approximately 1.5~eV above the valence band maximum set to 0~eV. This behavior is consistent with the linear one-photon absorption at $\hbar\omega_0=2.4$~eV. The conduction states are populated already when the pulse reaches its peak intensity ($t=18$~fs, Fig.~\ref{fig:occ_density}a) and remain partially occupied also after the laser is turned off ($t=36$~fs, Fig.~\ref{fig:occ_density}b). However, the post-pulse population of the conduction region increases roughly twice as much as that at 18 fs, indicating that most excited electrons remain there without significant recombination after the excitation process. Due to charge conservation, the population of the conduction states is combined with a corresponding depletion of the valence band maximum around -1~eV, see Fig.~\ref{fig:occ_density}a,b.

At the intermediate intensities of $400~\mathrm{GW/cm^2}$ and $800~\mathrm{GW/cm^2}$, when both $N_{ph}$ and $N_{ex}$ undergo saturation (Fig.~\ref{fig:E_N_both}), unoccupied states available at~3.9 eV are populated when the pulse reaches its peak ($t=18$~fs, Fig.~\ref{fig:occ_density}c), as a consequence of two-photon absorption. The overall occupation density in the conduction band becomes broader at $I=800~\mathrm{GW/cm^2}$, mirrored by a stronger depletion of the valence region. 
After the pulse is turned off at $t=36$~fs, the valence band population does not change further under either intensity. Still, there is a slight redistribution of the occupation in the conduction region, see Fig.~\ref{fig:occ_density}d. In particular, the population maximum 1.5~eV, associated with one-photon absorption, appears stronger for $I=400~\mathrm{GW/cm^2}$ than for $I=800~\mathrm{GW/cm^2}$, whereas at higher energies, the occupation density is larger when $I=800~\mathrm{GW/cm^2}$. This is consistent with the higher number of excited electrons but lower energy uptake at $I=400~\mathrm{GW/cm^2}$ (Fig.~\ref{fig:E_N_both}). At $I=800~\mathrm{GW/cm^2}$, additional population appears at 6.2~eV: this feature is ascribed to three-photon absorption. This result confirms the appearance of multi-photon absorption processes with intensities of $800~\mathrm{GW/cm^2}$ or higher.

The occupation density at $1~\mathrm{TW/cm^2}$ closely resembles the one at $800~\mathrm{GW/cm^2}$ with decreasing population at 1.5~eV (one-photon absorption), 3.9~eV (two-photon absorption), and 6.2~eV (three-photon absorption, see Fig.~\ref{fig:occ_density}e). Interestingly, also in this case, population of states at 6.2~eV appears only after laser illumination ($t=36$~fs, Fig.~\ref{fig:occ_density}f), suggesting that the excitation mechanisms at these intensities are equally dominated by multi-photon absorption. A considerable change in the laser-driven occupation of the conduction region occurs at $I=10~\mathrm{TW/cm^2}$. Under this intensity, the excited electron population is almost uniformly distributed up to 17.5~eV. This effect is more remarkable during irradiation (Fig.~\ref{fig:occ_density}e) than after the pulse is turned off (Fig.~\ref{fig:occ_density}f). The peaks at 1.5~eV, 3.9~eV, and 6.2~eV merge into an almost featureless continuous distribution that can be associated with emerging tunnel ionization. In this strong-field regime, tunneling through the distorted Coulomb potential barrier of the nuclei occurs~\cite{wang2012,jiao2013,wachter2014}. It is worth recalling that tunnel ionization depends on the field strength, not on the photon energy~\cite{Qi2024}, leading to a broad and continuous energy distribution of the excited electrons. This effect is also consistent with the loss of space inversion symmetry of the density distribution already discussed in the context of the HHG spectra (Fig.~\ref{fig:HHG}d). Finally, a slight increase in the occupation density around 12.5~eV can be related to higher-order nonlinear processes and other complex light-matter interactions occurring in this strong coupling regime. The valence band population decreases accordingly, halving its value compared to the result obtained under weak laser intensity (compare Fig.~\ref{fig:occ_density}a,b).


\section{Summary and Conclusions}
In summary, we investigated from first principles the time-dependent response of $\mathrm{Cs_3Sb}$ and $\mathrm{Cs_2Te}$ to ultrafast laser pulses of increasing intensities, ranging from $1~\mathrm{GW/cm^2}$ to $1~\mathrm{PW/cm^2}$, in order to explore different regimes of light-matter couplings. The carrier frequency and the polarization of the incoming field were set in resonance with the strongest linear absorption maxima of the two materials at 2.4~eV for cubic $\mathrm{Cs_3Sb}$ and 3.6~eV for orthorhombic $\mathrm{Cs_2Te}$, exciting this system with an $x$-polarized pulse. In this analysis, we monitored HHG, the number of absorbed photons and the number of excited electrons per unit cell, as well as the photon absorption efficiency. 

Up to $I=100~\mathrm{GW/cm^2}$, the response of the materials remains linear, as testified by the absence of high-harmonics paks in the HHG spectra and by linear increases of both $N_{ph}$ and $N_{ex}$ upon increasing laser intensities. For $100~\mathrm{GW/cm^2} \leq I \leq 1~\mathrm{TW/cm^2}$ in $\mathrm{Cs_3Sb}$ and $200~\mathrm{GW/cm^2} \leq I \leq 5~\mathrm{TW/cm^2}$ in $\mathrm{Cs_2Te}$, higher harmonic peaks emerge in the HHG spectra and the scaling of absorbed photons and excited electrons becomes nonlinear, indicating one-photon absorption saturation and the onset of nonlinear effects. 
For intensities above these ranges, the response of the materials becomes markedly nonlinear. The HHG spectra exhibit strong peaks at high harmonics and prominent signatures of symmetry breaking in the electronic distribution. The number of absorbed photons increases superlinearly with respect to the intensity, while the number of excited electrons grows more slowly due to Pauli blocking. Finally, the electron occupation density in the conduction region of $\mathrm{Cs_3Sb}$ becomes significantly broader and extends almost homogeneously up to high energies ($>10$~eV), suggesting the presence of multi-photon absorption and tunnel ionization effects.

In conclusion, this study provides crucial insights into the nonlinear optical properties of $\mathrm{Cs_3Sb}$ and $\mathrm{Cs_2Te}$ and highlights the complex interplay between intense laser radiation and the electronic structure of these materials. As such, these findings improve our general understanding of the response of Cs-based photocathode materials under intense laser fields. This knowledge is essential for optimizing their performance in particle accelerators, where they are subject to extreme irradiation to produce ultrabright electron beams.

\section*{Acknowledgements}
Useful discussions with Holger-Dietrich Sa{\ss}nick in the initial stage of this project are kindly acknowledged.
This work was funded by the German Research Foundation (DFG), Project No. 490940284, by the State of Lower Saxony (Professorinnen f\"ur Niedersachsen, DyNano, and ElLiKo), and by the German Federal Ministry of Education and Research (Professorinnenprogramm III). Computational resources were provided by the HPC cluster ROSA at the University of Oldenburg, funded by the DFG (project number INST 184/225-1 FUGG) and the Ministry of Science and Culture of the State of Lower Saxony.

\section*{Data availability}
The data that support the findings of this article are publicly available free of charge on Zenodo at the following link: https://doi.org/10.5281/zenodo.15005277


%

\end{document}